\documentclass{PoS}

\usepackage{amsmath,url}
\usepackage{graphicx}
\usepackage{cite}
\usepackage{subcaption}
\usepackage{multirow}

\newcommand{\citere}[1]{Ref.~\cite{#1}}
\newcommand{\citeres}[1]{Refs.~\cite{#1}}
\newcommand{\reffi}[1]{Fig.~\ref{#1}}

\newcommand{\refta}[1]{Tab.~\ref{#1}}

\newcommand\Matrix{{\sc Matrix}}
\newcommand\Munich{{\sc Munich}}
\newcommand\OpenLoops{{\sc OpenLoops}}
\newcommand\Collier{{\sc Collier}}
\newcommand{\CutTools}{{\sc CutTools}}
\newcommand{\OneLOop}{{\sc OneLOop}}

\newcommand{\abbrev}{}

\newcommand{\lo}{\text{\abbrev LO}}
\newcommand{\nlo}{\text{\abbrev NLO}}
\newcommand{\nnlo}{\text{\abbrev NNLO}}

\newcommand{\qT}{q_{\mathrm{T}}}

\newcommand{\qTVV}{q_{\mathrm{T}}^{\mathrm{VV}}}

\newcommand{\D}{\mathrm{d}}

\newcommand\Tstrut{\rule{0pt}{3.0ex}}         
\newcommand\Bstrut{\rule[-1.5ex]{0pt}{0pt}}   
\title{Status of diboson production in NNLO QCD}

\ShortTitle{Status of diboson production in NNLO QCD}

\author{Massimiliano Grazzini\\
        University of Zurich\\
        E-mail: \email{grazzini@physik.uzh.ch}}

\author{Stefan Kallweit\\
        Johannes-Gutenberg-University Mainz\\
        E-mail: \email{kallweit@uni-mainz.de}}

\author{\speaker{Dirk Rathlev}\\
        DESY Hamburg\\
        E-mail: \email{dirk.rathlev@desy.de}}

\abstract{We present results from the first fully differential next-to-next-to leading order NNLO QCD computation of ZZ production, including off-shell effects, the leptonic decay and non-resonant contributions. We compare theoretical predictions to fiducial cross sections and distributions measured by ATLAS and CMS at 8 TeV and also perform a first comparison with early ATLAS measurements at 13 TeV.}

\FullConference{12th International Symposium on Radiative Corrections (Radcor 2015) and LoopFest XIV (Radiative Corrections for the LHC and Future Colliders)\\
		 15-19 June  2015\\
		 UCLA Department of Physics and Astronomy Los Angeles, CA, USA}

\begin{document}

\section{Introduction}
ZZ production is an important Standard Model (SM) process for the LHC physics program. It provides a direct test of the electroweak (EW) sector of the SM. Furthermore, off-shell SM ZZ production is an irreducible background in Higgs searches and, in the high-mass region, for Higgs width measurements. Its high-mass tail is also sensitive to effects from anomalous couplings.

Until recently, the state-of-the-art fixed-order SM prediction for ZZ production, including the leptonic decay and off-shell effects, was the next-to-leading (NLO) QCD calculation from \citere{Campbell:2011bn}. In the following, we will present results from the first fully differential NNLO QCD computation of the process $pp\to 4\,\textrm{leptons}$, including spin correlations, off-shell effects and non-resonant contributions from $Z\gamma^*$ and $\gamma^*\gamma^*$ production. In the meantime some of these results have been published in \citere{Grazzini:2015hta}.

\section{Details of the calculation}
We have performed the NNLO calculation with the numerical program \Matrix\footnote{\Matrix{} is the abbreviation of ``\Munich{} Automates qT subtraction and Resummation to Integrate Cross Sections'', by M.~Grazzini, S.~Kallweit, D.~Rathlev, M.~Wiesemann. In preparation.}, which contains a process-independent implementation of the $\qT$ subtraction procedure \cite{Catani:2007vq}. In the \Matrix{} framework, \Munich \footnote{\Munich{} is the abbreviation of ``MUlti-chaNnel Integrator at Swiss~(CH) precision''---an automated parton level NLO, by S.~Kallweit. In preparation.} takes care of the phase-space integration, the construction of the necessary Catani-Seymour (CS) dipoles \cite{Catani:1996jh,Catani:1996vz} and also provides an interface to the one-loop generator \OpenLoops{} \cite{Cascioli:2011va}. \OpenLoops{}, together with the \Collier{} library~\cite{Denner:2014gla,Denner:2002ii,Denner:2005nn,Denner:2010tr}, is used for the evaluation of all tree-level and one-loop amplitudes. To deal with problematic phase-space points, \OpenLoops{} provides a rescue system,
which uses the quadruple-precision implementation of the OPP method in 
\CutTools{}~\cite{Ossola:2007ax} and scalar integrals from 
\OneLOop{}~\cite{vanHameren:2010cp}.

\Matrix{} has also been used in the NNLO computations of \citeres{Grazzini:2013bna,Cascioli:2014yka,Gehrmann:2014fva,Grazzini:2015nwa}, and in the 
resummed calculation of \citere{Grazzini:2015wpa}.

For the handling of infrared singularities at NNLO we use the $\qT$ subtraction formalism. $\qT$ subtraction renders the separation between genuine NNLO singularities, characterised by the limit in which the transverse momentum of the diboson pair, $\qTVV$, approaches zero, from NLO-like singularities in the $\textrm{VV+jet}$ contribution fully transparent. This implies that the real contribution $\D{\sigma}^{\mathrm{VV + jet}}$ in its master formula,
 \begin{align}
\label{eq:main}
\D{\sigma}^{\mathrm{VV}}_{\mathrm{NNLO}}={\cal H}^{\mathrm{VV}}_{\mathrm{NNLO}}\otimes \D{\sigma}^{\mathrm{VV}}_{\mathrm{LO}}+\left[ \D{\sigma}^{\mathrm{VV + jet}}_{\mathrm{NLO}}-\D{\sigma}^{\mathrm{CT}}_{\mathrm{NNLO}}\right],
\end{align}
can be evaluated using any NLO subtraction procedure. The divergence of the real contribution in the limit $\qTVV\to0$ is cancelled by a process-independent counterterm $\D{\sigma}^{\mathrm{CT}}$. The one- and two-loop virtual corrections, which live on the Born phase-space,  enter via the hard function, ${\cal H}^{\mathrm{VV}}$.

\begin{figure}[h]
  \begin{subfigure}[b]{0.48\linewidth}
    \centering
    \includegraphics[scale=0.6]{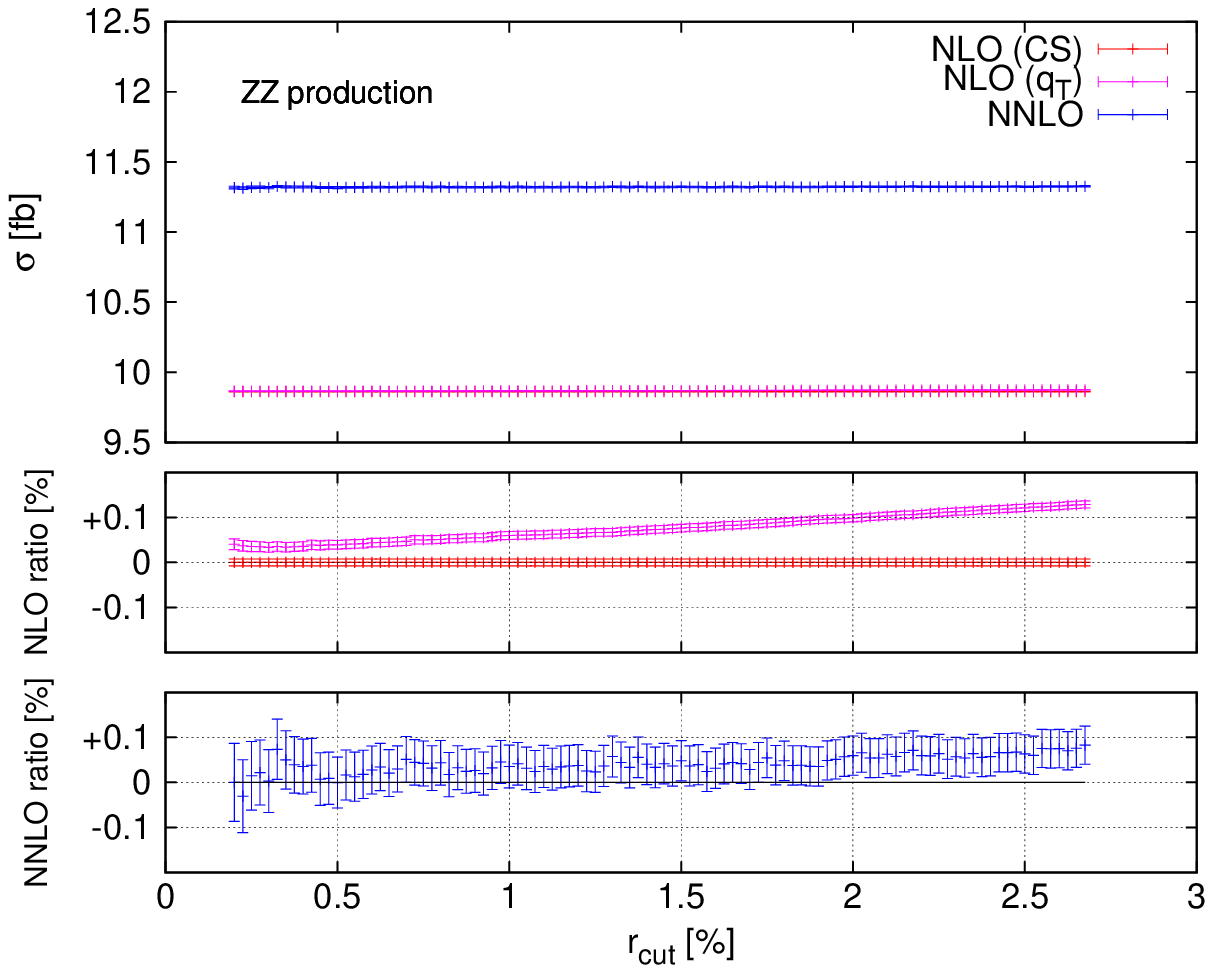} 
    \caption{$ZZ$ production.} 
  \end{subfigure} 
  \begin{subfigure}[b]{0.48\linewidth}
    \centering
    \includegraphics[scale=0.6]{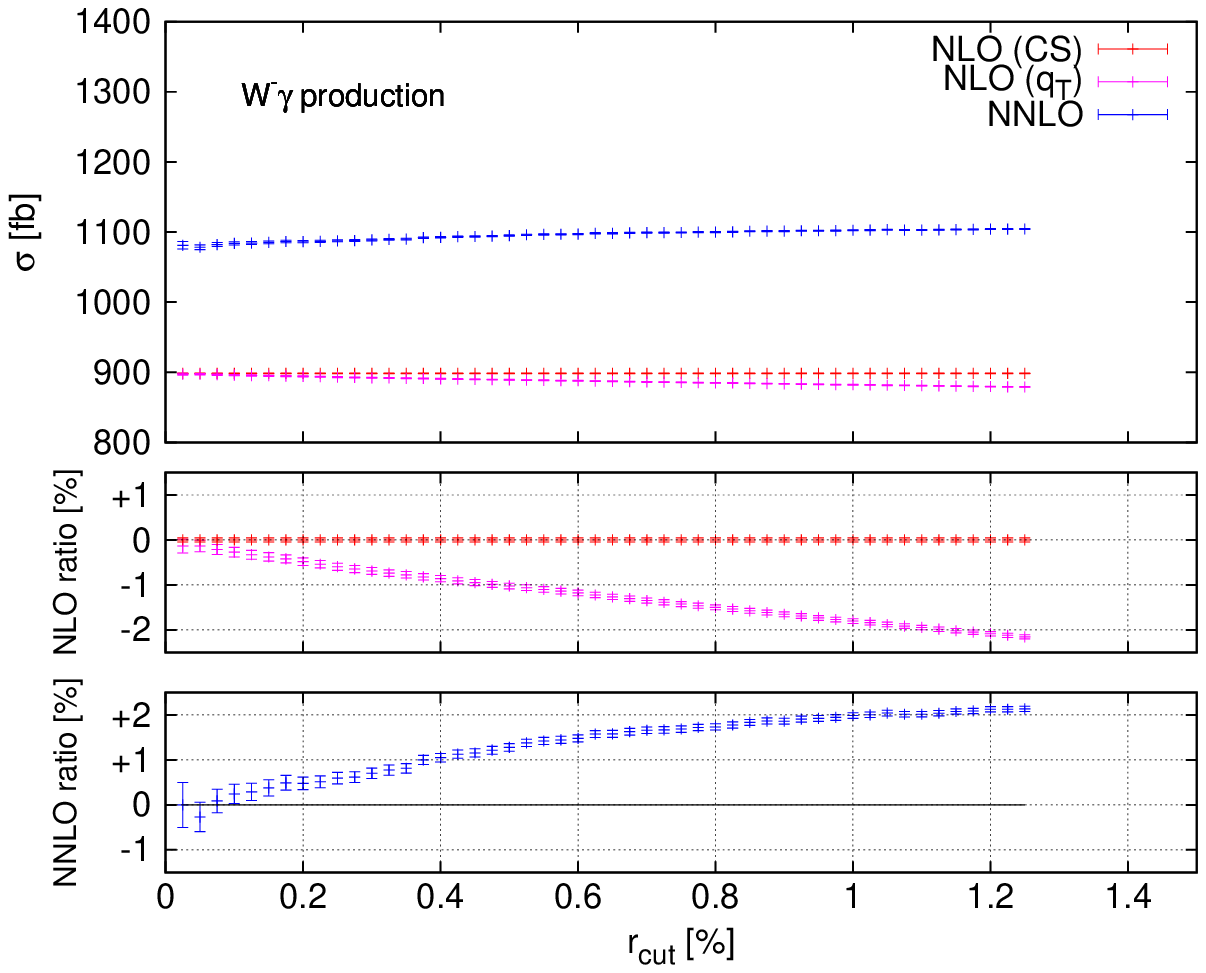} 
    \caption{$W^-\gamma$ production.} 
  \end{subfigure} 
  \caption{$r_{\mathrm{cut}}$ dependence of the NLO (magenta) and NNLO (blue) cross section for $ZZ$ and $W^-\gamma$ production. The $r_{\mathrm{cut}}$ independent NLO cross section computed with CS subtraction is also shown (red). The lower panels show the ratio of the NLO cross section computed with $\qT$ subtraction over the same NLO cross section computed with CS subtraction, and the ratio of the NNLO cross section over the NNLO cross section evaluated at the smallest considered value of $r_{\mathrm{cut}}$.}
  \label{fig:qtvars}
\end{figure}

$\qT$ subtraction is a non-local subtraction method w.r.t.\ the $\qT\to0$ singularity, and in practical implementations a small technical cut $r_{\mathrm{cut}}$ needs to be applied on $r\equiv \qT/Q$, where $Q$ is the invariant mass of the final-state system, in this case $Q=m_{\mathrm{VV}}$. For sufficiently small values of $r_{\mathrm{cut}}$ the cross section should become cut-independent. The qualitiy of the cancellation between real contribution and counterterm, and the optimal size of the technical cut depend on the process. \reffi{fig:qtvars} shows the cut dependence of the total cross section for two benchmark processes, $ZZ$ production (left) and $W\gamma$ production (right). In the case of $ZZ$ production, the result is very stable when varying the cut, and even relatively large values of $r_{\mathrm{cut}}\approx 10^{-2}$ give reliable results, while in the case of $W\gamma$ production the cut dependence is significantly stronger and very small cut values below $r_{\mathrm{cut}}\approx 10^{-3}$ are required. The origin of the stronger cut dependence in the case of $W\gamma$ production lies in the photon isolation. More details can be found in \citere{Grazzini:2015nwa}.

\section{Numerical results}
As a benchmark, we consider $pp$ collisions at $\sqrt{s}=8$ TeV. We use the so-called $G_\mu$ scheme, in which the input parameters are $G_F=1.16639\times 10^{-5}$~GeV$^{-2}$, $m_W=80.399$ GeV and
$m_Z = 91.1876$~GeV, and $\cos\theta_W$ and $\alpha_{EW}$ are computed from these. Consistently with \OpenLoops, we use the complex mass scheme for the $W$ and $Z$ bosons, in which the weak mixing angle is defined via $\cos\theta_W^2=(m_W^2-i\Gamma_W\,m_W)/(m_Z^2-i\Gamma_Z\,m_Z)$, and we set $\Gamma_W=2.1054$ GeV and $\Gamma_Z=2.4952$ GeV. Contributions with closed top-quark loops in the real-virtual and gluon-fusion contribution are also sensitive to the masses and widths of the top quark and the Higgs boson, which we set to $m_t=173.2$ GeV and $\Gamma_t=1.4426$ GeV, and to $m_H=125$ GeV and $\Gamma_H=4.07$ MeV, respectively. We employ NNPDF3.0 PDF sets~\cite{Ball:2014uwa}, where we use consistent sets and $\alpha_S$ running at each perturbative order, i.e.\ NLO sets and two-loop running at NLO and NNLO sets and three-loop running at NNLO. We consider $N_f=5$ massless quark flavors in the initial and final states. The selection cuts we employ are quite inclusive, justifying fixed renormalization and factorization scales $\mu_R=\mu_F=m_Z$. We estimate perturbative uncertainties due to missing higher-order corrections by independently varying both scales up and down by a factor of two, where we exclude the antipodal variations to avoid large logarithms of $\mu_R/\mu_F$.

We consider two benchmark scenarios. The first is given by the ATLAS ZZ analysis at 8 TeV presented in \citere{ATLAS-CONF-2013-020}. The three
decay channels $e^+e^-e^+e^-$, $\mu^+\mu^-\mu^+\mu^-$, and $e^+e^-\mu^+\mu^-$ are considered separately. The ATLAS analysis employs an invariant mass cut of 66~GeV $\leq m_{ll}\leq$~116 GeV on the reconstructed Z bosons. The pairing ambiguity in the equal-flavor cases is resolved by choosing the pairing which minimizes the sum of the absolute differences between the reconstructed invariant masses and the physical Z mass. The lepton cuts do not discriminate between electrons and muons and read 
$p_T\geq 7$~GeV and $|\eta|\leq 2.7$. The lepton isolation requirement is given by $\Delta R(l,l^\prime)>0.2$ for any lepton pair in the final state. Note that this cut is necessary in the equal-flavor case to obtain an infrared safe cross section definition.

The predicted fiducial cross sections and the measured cross sections from ATLAS are reported in \refta{tableatlas}. Consistently with the size of the higher-order corrections in the case of on-shell ZZ production \cite{Cascioli:2014yka}, the NNLO effects amount to a correction of about 15\% compared to the NLO cross sections. The gluon-fusion channel opening up at $\mathcal{O}(\alpha_S^2)$ contributes around 60\% of the NNLO corrections, the rest coming from corrections to the $q\bar{q}$ channel. The scale uncertainties, which stay at the $\pm3\%$ level at NNLO, are also dominated by the gluon-fusion contribution. We note that in the meantime, first results for the NLO corrections to the gluon-fusion channel have appeared in \citere{Caola:2015psa}, indicating sizeable corrections to this channel at $\mathcal{O}(\alpha_S^3)$.

Comparing with the experimentally measured cross sections from ATLAS, we note that the inclusion of the full NNLO corrections improves the agreement with data slightly in the different-flavor case, but leads to predicted cross sections that slightly overshoot the data in the same-flavor channels. We note, however, that the experimental uncertainties are still relatively large and all NNLO predictions are consistent with data at the $1\sigma$ level.

\renewcommand{\baselinestretch}{1.5}
\begin{table}[ht]
\begin{center}
\begin{tabular}{|c| c| c| c| c|}
\hline
Channel & $\sigma_{\textrm{LO}}$ (fb) & $\sigma_{\textrm{NLO}}$ (fb) & $\sigma_{\textrm{NNLO}}$ (fb) & $\sigma_{\textrm{exp}}$ (fb) \\ [0.5ex]
\hline
\Tstrut
$e^+e^-e^+e^-$ & \multirow{2}{*}{$3.547(1)^{+2.9\%}_{-3.9\%}$} & \multirow{2}{*}{$5.047(1)^{+2.8\%}_{-2.3\%}$} & \multirow{2}{*}{$5.79(2)^{+3.4\%}_{-2.6\%}$} & $4.6^{+0.8}_{-0.7}{\rm (stat)}^{+0.4}_{-0.4}{\rm (syst.)}^{+0.1}_{-0.1}{\rm (lumi.)}$\Bstrut\\
\cline{1-1}
\cline{5-5}
\Tstrut
$\mu^+\mu^-\mu^+\mu^-$ &  &  &  & $5.0^{+0.6}_{-0.5}{\rm (stat)}^{+0.2}_{-0.2}{\rm (syst.)}^{+0.2}_{-0.2}{\rm (lumi.)}$\Bstrut\\
\hline
\Tstrut
 $e^+e^-\mu^+\mu^-$ & $6.950(1)^{+2.9\%}_{-3.9\%}$ & $9.864(2)^{+2.8\%}_{-2.3\%}$ & $11.31(2)^{+3.2\%}_{-2.5\%}$ & $11.1^{+1.0}_{-0.9}{\rm (stat)}^{+0.5}_{-0.5}{\rm (syst.)}^{+0.3}_{-0.3}{\rm (lumi.)}$\Bstrut\\
\hline
\end{tabular}
\end{center}
\renewcommand{\baselinestretch}{1.0}
\caption{\label{tableatlas} Fiducial cross sections and scale uncertainties at 8 TeV for ATLAS cuts 
at \lo{}, \nlo{}, and \nnlo{} in the three considered leptonic decay channels. The ATLAS data are also shown.}
\end{table}

In the meantime, a first analysis of ZZ production at 13 TeV has been presented by ATLAS in \citere{Aad:2015zqe}. The fiducial cuts are the same as in the 8 TeV analysis, except for the minimum lepton transverse momentum, which has been increased to $p_T\geq 20$~GeV. \refta{tableatlas13} shows the theoretical predictions at 13 TeV, for which now CT10 PDF sets \cite{Gao:2013xoa} and a dynamical scale $\mu_R=\mu_F=m_{ZZ}/2$ have been used. At the moment, the experimental precision is severely limited by statistics. However, in general the NNLO predictions are in very good agreement with data.

\renewcommand{\baselinestretch}{1.5}
\begin{table}[ht]
\begin{center}
\begin{tabular}{|c| c| c| c| c|}
\hline
Channel & $\sigma_{\textrm{LO}}$ (fb) & $\sigma_{\textrm{NLO}}$ (fb) & $\sigma_{\textrm{NNLO}}$ (fb) & $\sigma_{\textrm{exp}}$ (fb) \\ [0.5ex]
\hline
\Tstrut
$e^+e^-e^+e^-$ & \multirow{2}{*}{$5.007(1)^{+4\%}_{-5\%}$} & \multirow{2}{*}{$6.157(1)^{+2\%}_{-2\%}$} & \multirow{2}{*}{$7.14(2)^{+2\%}_{-2\%}$} & $8.4^{+2.4}_{-2.0}{\rm (stat)}^{+0.4}_{-0.2}{\rm (syst.)}^{+0.5}_{-0.3}{\rm (lumi.)}$\Bstrut\\
\cline{1-1}
\cline{5-5}
\Tstrut
$\mu^+\mu^-\mu^+\mu^-$ &  &  &  & $6.8^{+1.8}_{-1.5}{\rm (stat)}^{+0.3}_{-0.3}{\rm (syst.)}^{+0.4}_{-0.3}{\rm (lumi.)}$\Bstrut\\
\hline
\Tstrut
 $e^+e^-\mu^+\mu^-$ & $9.906(1)^{+4\%}_{-5\%}$ & $12.171(2)^{+2\%}_{-2\%}$ & $14.19(2)^{+2\%}_{-2\%}$ & $14.7^{+2.9}_{-2.5}{\rm (stat)}^{+0.6}_{-0.4}{\rm (syst.)}^{+0.9}_{-0.6}{\rm (lumi.)}$\Bstrut\\
\hline
\end{tabular}
\end{center}
\renewcommand{\baselinestretch}{1.0}
\caption{\label{tableatlas13} Fiducial cross sections and scale uncertainties at 13 TeV for ATLAS cuts 
at \lo{}, \nlo{}, and \nnlo{} in the three considered leptonic decay channels. The ATLAS data are also shown.}
\end{table}

We now move on to our second benchmark setup, based on the CMS analysis at 8 TeV in \citere{CMS:2014xja}. The fiducial cuts used by CMS differentiate between electrons and muons and read as follows: the muons are required to fulfill $p_T^\mu>5$ GeV, $|\eta^\mu|<2.4$, while the electrons are required to fulfill $p_T^e>7$~GeV, $|\eta^e|<2.5$. In addition, the leading- 
and subleading-lepton transverse momenta must satisfy $p_T^{l,1}>20$ GeV and $p_T^{l,2}>10$ GeV, 
respectively. The pairing ambiguity is resolved by choosing the same-flavor opposite-sign lepton pair whose invariant mass is closest to the Z mass as the first, and the remaining pair as the second reconstructed Z boson. The invariant masses of both reconstructed Z bosons are required to satisfy 60 GeV $\leq m_{ll}\leq$ 120 GeV. In the case of equal-flavor leptons in the final state, an additional cut is needed to render the fiducial cross section infrared finite. Instead of the $\Delta R(l,l^\prime)$ cut used by ATLAS, CMS employs a lower cut on the invariant mass of any lepton pair in the final state, $ m_{ll}>4$ GeV.

We compute the theoretical uncertainties as above. The predicted fiducial cross sections are reported in \refta{tablecms}. We note that the relative impact of the NNLO corrections is very similar to the one found with ATLAS cuts. Also the scale uncertainties are very similar to those reported in \refta{tableatlas}.

\renewcommand{\baselinestretch}{1.5}
\begin{table}[ht]
\begin{center}
\begin{tabular}{|c| c| c| c|}
\hline
Channel & $\sigma_{\textrm{LO}}$ (fb) & $\sigma_{\textrm{NLO}}$ (fb) & $\sigma_{\textrm{NNLO}}$ (fb) \\ [0.5ex]
\hline
\Tstrut
$e^+e^-e^+e^-$ & $3.149(1)^{+3.0\%}_{-4.0\%}$ & $4.493(1)^{+2.8\%}_{-2.3\%}$ & $5.16(1)^{+3.3\%}_{-2.6\%}$ \Bstrut\\
\hline
\Tstrut
$\mu^+\mu^-\mu^+\mu^-$ & $2.973(1)^{+3.1\%}_{-4.1\%}$ & $4.255(1)^{+2.8\%}_{-2.3\%}$ & $4.90(1)^{+3.4\%}_{-2.6\%}$\Bstrut\\
\hline
\Tstrut
 $e^+e^-\mu^+\mu^-$ & $6.179(1)^{+3.1\%}_{-4.0\%}$ & $8.822(1)^{+2.8\%}_{-2.3\%}$ & $10.15(2)^{+3.3\%}_{-2.6\%}$ \Bstrut\\
\hline
\end{tabular}
\end{center}
\renewcommand{\baselinestretch}{1.0}
\caption{\label{tablecms}
Fiducial cross sections and scale uncertainties at 8 TeV for CMS cuts at \lo{}, \nlo{}, and \nnlo{} in the 
three considered leptonic decay channels.}
\end{table}

\begin{figure}[htpb]
        \centering
        \includegraphics[width=0.8\textwidth]{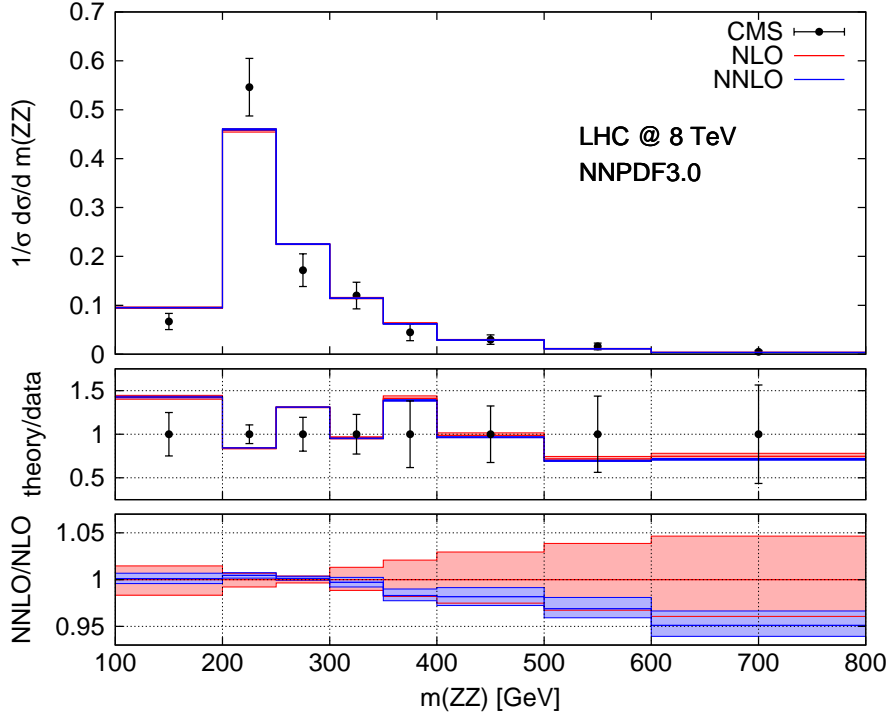}
    \caption{The four-lepton invariant-mass distribution
    at \nlo{} and \nnlo{} compared to the CMS data. In the lower panels the ratio of our theoretical results 
over the data, and the \nnlo{} result normalized to the central \nlo{} prediction are presented. The bands correspond to scale variations as described in the text.}
    \label{fig:1}
\end{figure}

While the CMS analysis does not report the measured fiducial cross sections, it does provide a number of normalized distributions, with which we can compare our theoretical predictions. \reffi{fig:1} shows the invariant-mass distribution of the four-lepton system. While the agreement between data and theory is generally good, the experimental uncertainties are still relatively large. In addition, normalizing the distribution by the fiducial cross section cancels out a significant part of the NNLO corrections, in particular in the peak region, where the cross section is measured most precisely. The lower panel in \reffi{fig:1} shows the ratio of the NNLO and the NLO prediction and indicates that the NNLO corrections have the effect of making the invariant mass distribution slightly softer. This can be traced back to the gluon-fusion contribution, which drops off more quickly at high invariant masses, where larger values of Bjorken $x$ are probed.

\begin{figure}[htpb]
    \begin{subfigure}[b]{0.49\textwidth}
        \centering
        \includegraphics[width=\textwidth]{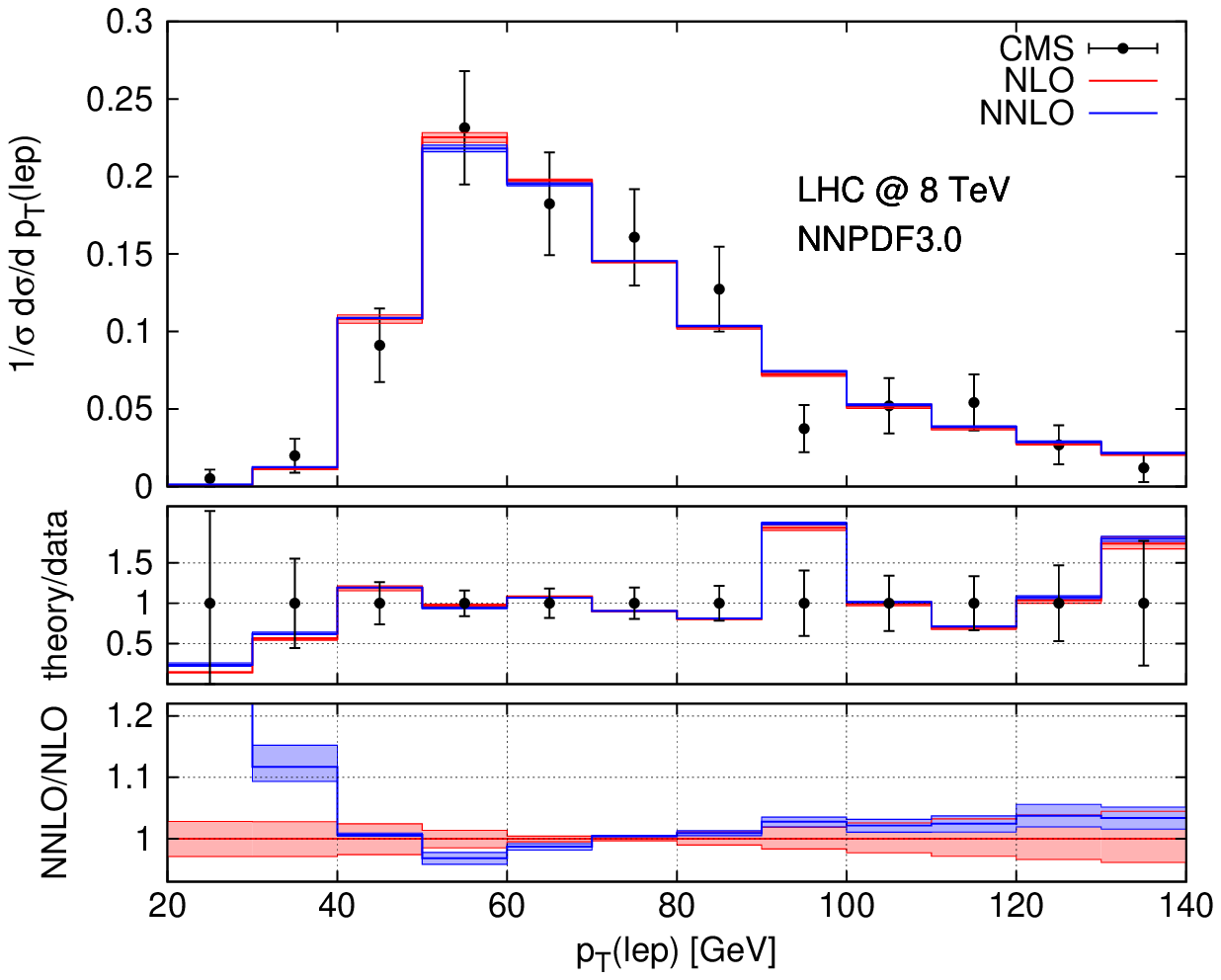}
    \end{subfigure}
    \begin{subfigure}[b]{0.49\textwidth}
        \centering
        \includegraphics[width=\textwidth]{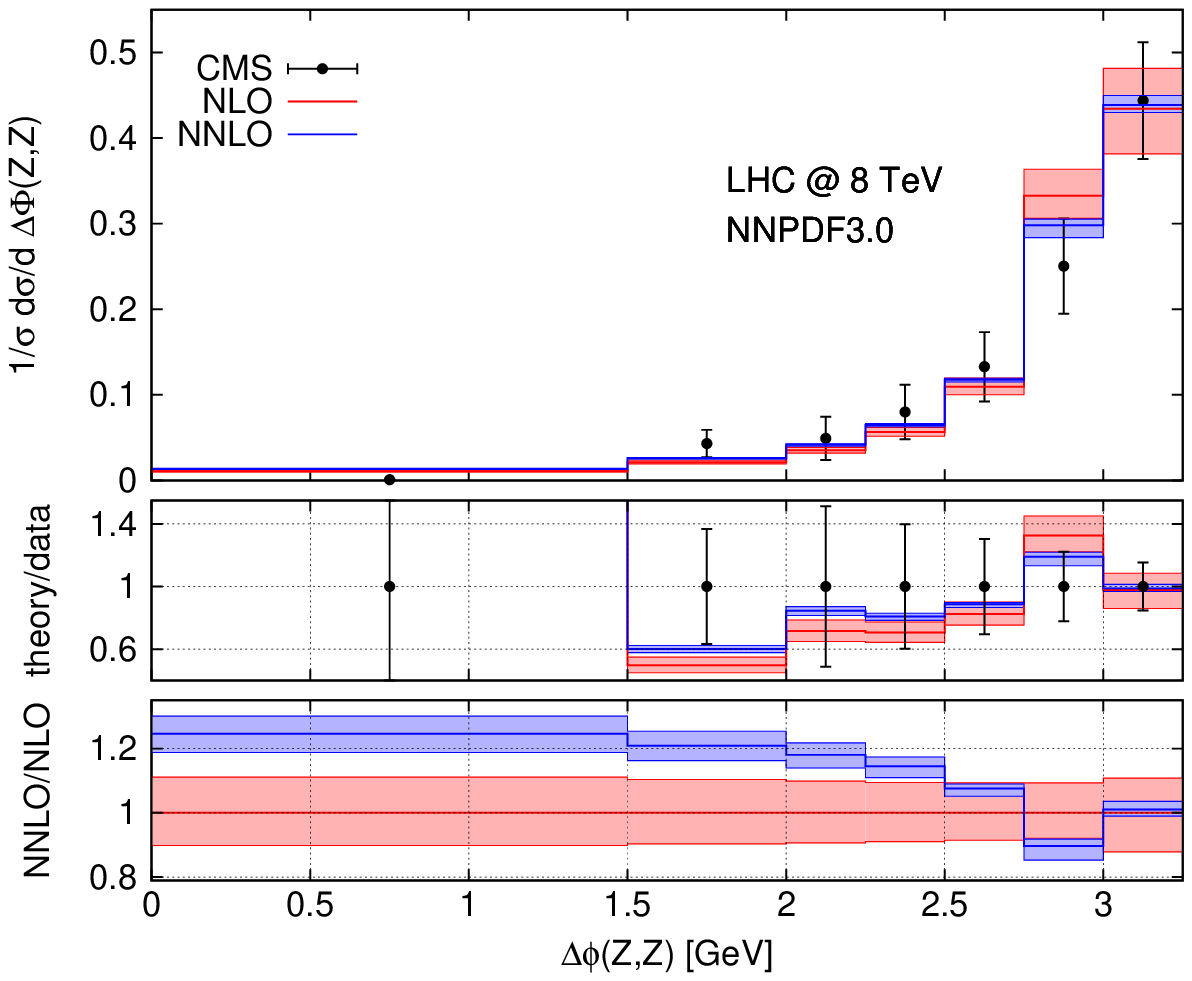}
    \end{subfigure}
    \caption{The leading-lepton $p_T$ (left) and the $\Delta\phi$ (right) distributions
    at \nlo{} and \nnlo{} compared to the CMS data.
In the lower panels the ratio of our theoretical results over the data, and the \nnlo{} result normalized to the central \nlo{} prediction are presented. The bands correspond to scale variations as described in the text.}
    \label{fig:2}
\end{figure}

\reffi{fig:2} shows the analogous results for the leading-lepton $p_T$ (left) and the azimuthal separation $\Delta\Phi$ between the two reconstructed Z bosons (right). Similar to the invariant-mass spectrum, the agreement between data and theory is good in the case of the lepton $p_T$, but the impact of NNLO effects is largely cancelled out by the normalization. The NNLO corrections are enhanced in the low-$p_T$ range, which is again due to the gluon-fusion contribution entering at $\mathcal{O}(\alpha_S^2)$.

The $\Delta\Phi$ distribution shows much larger NNLO effects, even when normalized to the fiducial cross section. This can largely be understood by the observation that at LO the Z bosons are always back-to-back and thus $\Delta\Phi=\pi$. For the $\Delta\Phi$ distribution the NLO is thus the first non-trivial order and the NNLO corrections are de-facto of NLO importance. We note, however, that the full NNLO cross section enters in the normalization, and thus the result shown in \reffi{fig:2} is a genuine NNLO prediction.

\section{Summary and discussion}
We reported on the first fully differential computation of NNLO QCD corrections to the production of 4 leptons at the LHC. While the relative size of the NNLO effects is similar to the one found in the on-shell computation, taking off-shell effects and the decay into account allowed for the first time to apply realistic selection cuts and to perform a direct comparison with measured fiducial cross sections and distributions.

The present study represents one of the first applications of the numerical program \Matrix{}, which is able to compute NNLO QCD corrections and to perform transverse-momentum resummation up to next-to-next-to-leading logarithmic accuracy for a wide class of processes relevant at the LHC.

\vspace{1cm}
\noindent {\bf Acknowledgements.}
This research was supported in part by the Swiss National Science Foundation (SNF) under 
contracts CRSII2-141847, 200021-156585, and by 
the Research Executive Agency (REA) of the European Union under the Grant Agreement 
number PITN--GA--2012--316704 ({\it HiggsTools}).


\begin{thebibliography}{99}
\bibitem{Campbell:2011bn}
  J.~M.~Campbell, R.~K.~Ellis and C.~Williams,
  JHEP {\bf 1107} (2011) 018
  [arXiv:1105.0020 [hep-ph]].

\bibitem{Grazzini:2015hta}
  M.~Grazzini, S.~Kallweit and D.~Rathlev,
  Phys.\ Lett.\ B {\bf 750} (2015) 407
  [arXiv:1507.06257 [hep-ph]].

\bibitem{Catani:2007vq}
  S.~Catani and M.~Grazzini,
  Phys.\ Rev.\ Lett.\  {\bf 98} (2007) 222002
[hep-ph/0703012].

\bibitem{Catani:1996jh}
  S.~Catani and M.~H.~Seymour,
  Phys.\ Lett.\ B {\bf 378} (1996) 287
  [hep-ph/9602277].

\bibitem{Catani:1996vz}
  S.~Catani and M.~H.~Seymour,
  Nucl.\ Phys.\ B {\bf 485} (1997) 291
   [Erratum-ibid.\ B {\bf 510} (1998) 503]
[hep-ph/9605323].


\bibitem{Cascioli:2011va}
  F.~Cascioli, P.~Maierh\"ofer and S.~Pozzorini,
  Phys.\ Rev.\ Lett.\  {\bf 108} (2012) 111601
  [arXiv:1111.5206 [hep-ph]].

\bibitem{Denner:2014gla}
  A.~Denner, S.~Dittmaier and L.~Hofer,
  PoS LL {\bf 2014} (2014) 071
  [arXiv:1407.0087 [hep-ph]].



\bibitem{Denner:2002ii}
  A.~Denner and S.~Dittmaier,
  Nucl.\ Phys.\ B {\bf 658} (2003) 175
  [hep-ph/0212259].


\bibitem{Denner:2005nn}
  A.~Denner and S.~Dittmaier,
  Nucl.\ Phys.\ B {\bf 734} (2006) 62
  [hep-ph/0509141].

\bibitem{Denner:2010tr}
  A.~Denner and S.~Dittmaier,
  Nucl.\ Phys.\ B {\bf 844} (2011) 199
  [arXiv:1005.2076 [hep-ph]].

\bibitem{Ossola:2007ax}
  G.~Ossola, C.~G.~Papadopoulos and R.~Pittau,
  JHEP {\bf 0803} (2008) 042
  [arXiv:0711.3596 [hep-ph]].

\bibitem{vanHameren:2010cp}
  A.~van Hameren,
  Comput.\ Phys.\ Commun.\  {\bf 182} (2011) 2427
  [arXiv:1007.4716 [hep-ph]].


\bibitem{Cascioli:2014yka}
  F.~Cascioli, T.~Gehrmann, M.~Grazzini, S.~Kallweit, P.~Maierh\"ofer, A.~von Manteuffel, S.~Pozzorini, D.~Rathlev, L.~Tancredi and E.~Weihs,
  Phys.\ Lett.\ B {\bf 735} (2014) 311
  [arXiv:1405.2219 [hep-ph]].


\bibitem{Grazzini:2013bna}
  M.~Grazzini, S.~Kallweit, D.~Rathlev and A.~Torre,
  Phys.\ Lett.\ B {\bf 731} (2014) 204
  [arXiv:1309.7000 [hep-ph]].


\bibitem{Gehrmann:2014fva}
  T.~Gehrmann, M.~Grazzini, S.~Kallweit, P.~Maierh\"ofer, A.~von Manteuffel, S.~Pozzorini, D.~Rathlev and L.~Tancredi,
  Phys.\ Rev.\ Lett.\  {\bf 113} (2014) 21,  212001
  [arXiv:1408.5243 [hep-ph]].

\bibitem{Grazzini:2015nwa}
  M.~Grazzini, S.~Kallweit and D.~Rathlev,
  arXiv:1504.01330 [hep-ph].

\bibitem{Grazzini:2015wpa}
  M.~Grazzini, S.~Kallweit, D.~Rathlev and M.~Wiesemann,
  arXiv:1507.02565 [hep-ph].


\bibitem{Ball:2014uwa}
  R.~D.~Ball {\it et al.}  [NNPDF Collaboration],
  JHEP {\bf 1504} (2015) 040
  [arXiv:1410.8849 [hep-ph]].


\bibitem{ATLAS-CONF-2013-020}
ATLAS Collaboration, ATLAS-CONF-2013-020.

\bibitem{Caola:2015psa}
  F.~Caola, K.~Melnikov, R.~Röntsch and L.~Tancredi,
  Phys.\ Rev.\ D {\bf 92} (2015) 9,  094028
  doi:10.1103/PhysRevD.92.094028
  [arXiv:1509.06734 [hep-ph]].

\bibitem{Aad:2015zqe}
  G.~Aad {\it et al.} [ATLAS Collaboration],
  arXiv:1512.05314 [hep-ex].

\bibitem{Gao:2013xoa}
  J.~Gao {\it et al.},
  Phys.\ Rev.\ D {\bf 89} (2014) 3,  033009
  doi:10.1103/PhysRevD.89.033009
  [arXiv:1302.6246 [hep-ph]].

\bibitem{CMS:2014xja}
  V.~Khachatryan {\it et al.}  [CMS Collaboration],
  Phys.\ Lett.\ B {\bf 740} (2015) 250
  [arXiv:1406.0113 [hep-ex]].


\end{thebibliography}
\end{document}